\documentclass{elsart41}
\usepackage{graphics}
\usepackage{graphicx}
\usepackage{amssymb}

\begin{document}

\begin{frontmatter}



\title{Large mass enhancement in RbOs$_2$O$_6$}
%

\author[CH]{M.~Br\"uhwiler\corauthref{cor1}}
\ead{markus.bruehwiler@phys.ethz.ch}
\author[CH]{S.M.~Kazakov}
\author[CH]{J.~Karpinski}
\author[CH]{B.~Batlogg}

\address[CH]{Laboratory for Solid State Physics, ETH Z\"urich, 8093 Z\"urich, Switzerland}

\corauth[cor1]{Corresponding author. Tel: +41 1 633 23 39 fax: +41
1 633 10 72}

\newcommand{\fu}{$\mathrm{RbOs_2O_6}$}
\newcommand*{\unit}[1]{\,\mathrm{#1}}
\newcommand{\ce}{\Delta U}  
\newcommand{\CE}{\widetilde{\ce}}  
\newcommand{\sfc}{\gamma}  
\newcommand{\SFC}{\widetilde{\sfc}}  
\newcommand{\Tc}{T_\mathrm{c}}  
\newcommand{\Cp}{C_\mathrm{p}}  
\newcommand{\Hc}{H_\mathrm{c}}  
\newcommand{\svf}{superconducting volume fraction}
\newcommand{\smf}{superconducting mass fraction}

\begin{abstract}

Heat capacity measurements on the recently discovered
geometrically frustrated $\beta$-pyrochlore superconductor \fu\
($\Tc=6.4\unit{K}$) yield a Sommerfeld coefficient of
$44\unit{mJ/mol_\mathrm{f.u.}/K^2}$. This is about $4$ times
larger than the one found in band structure calculations. In order
to specify the enhancement due to electron-electron interactions,
we have measured the electron-phonon enhancement. By a suitable
analysis, an electron-phonon coupling constant
$\lambda_\mathrm{ep} = 1\pm0.1$ is derived from the specific heat
jump at $\Tc$. This leaves a significant additional
$\lambda_\mathrm{add} = 2.1\pm0.3$ for enhancement due to other
mechanisms, possibly related to the triangular lattice. To arrive
at these results, an appropriate analysis method for bulk
thermodynamic data based on the condensation energy was applied.

\end{abstract}

\end{frontmatter}

\newcommand{\fu}{$\mathrm{RbOs_2O_6}$}
\newcommand*{\unit}[1]{\,\mathrm{#1}}
\newcommand{\ce}{\Delta U}  
\newcommand{\CE}{\widetilde{\ce}}  
\newcommand{\sfc}{\gamma}  
\newcommand{\SFC}{\widetilde{\sfc}}  
\newcommand{\Tc}{T_\mathrm{c}}  
\newcommand{\Cp}{C_\mathrm{p}}  
\newcommand{\Hc}{H_\mathrm{c}}  
\newcommand{\svf}{superconducting volume fraction}
\newcommand{\smf}{superconducting mass fraction}


Long standing interest in the pyrochlores stems from their
inherent geometrical frustration due to the metal ions forming a
network of corner-sharing tetrahedra. Thus, metallic pyrochlores
constitute ideal systems to study to what degree itinerant
electrons are affected by a lattice which is known to cause
geometrical frustration for interactions among localized magnetic
moments. For similar reasons, the superconductivity recently found
in \fu\ has been of considerable interest. \fu\ is one of only
four pyrochlore superconductors known to date. These are the
$\alpha$-pyrochlore Cd$_2$Re$_2$O$_7$ and the $\beta$-pyrochlores
\textit{A}Os$_2$O$_6$, where \textit{A} = Cs, Rb, or
K.\cite{YoMuHi2004,YoMuMaHi2004a,YoMuMaHi2004} \fu\ is a
conventional $s$-wave superconductor with a critical temperature
$\Tc =
6.4\unit{K}$.\cite{BrKaZhKaBa2004,KhEsKaKaZhBrGaDiShMaMaBaKe2004,MaGaPeHiWeOtKaKa2004,BrKaKaBa2005}
In this short paper, we provide evidence for an additional
electronic mass enhancement beyond the contribution from the
coupling to phonons. Furthermore, the basic thermodynamic
parameters of \fu\ are extracted.

Specific heat measurements on \fu\ show that the residual
Sommerfeld coefficient in the superconducting state
$\sfc_\mathrm{r}$ and the normal-state coefficient $\sfc$ vary
among the various samples measured. The variation is consistent
with the presence of a second metallic component, and subsequent
x-ray diffraction analysis has confirmed the presence of OsO$_2$
in the samples. We have therefore developed a suitable
quantitative method to analyze thermodynamic data when dealing
with a superconducting sample containing a metallic second phase.
Since the analysis is based on the condensation energy of the
superconductor of interest, we call it condensation energy
analysis (CEA). It involves integrating the heat capacity to
obtain the condensation energy which is a reliable measure of the
superconducting fraction. From the systematic variation of the
thermodynamic parameters on the superconducting fraction it is
possible to extract the properties of the pure
sample.\cite{BrKaKaBa2005}

\begin{figure}[!ht]
\begin{center}
\includegraphics[width=0.45\textwidth]{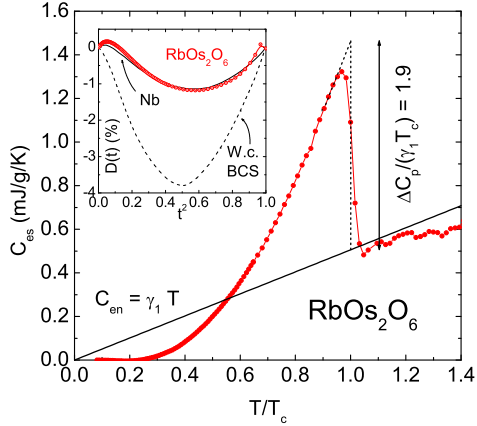}
\end{center}
\caption{\label{fig:Ces_vs_TdivTc} Superconducting electronic
specific heat of \fu. The normalized specific heat jump $\Delta
\Cp\vert_{T_\mathrm{c}}/(\sfc_1 \Tc) = 1.9$ is significantly
larger than that for a weak coupling superconductor and
corresponds to an electron-phonon coupling constant
$\lambda_\mathrm{ep} = 1\pm0.1$. The inset shows the deviation of
the normalized critical field $\Hc/\Hc(0)$ from a simple $1-t^2$
behavior.}
\end{figure}

According to the CEA, the superconducting electronic specific heat
$C_\mathrm{es}$ is extracted from the measurement by
$C_\mathrm{es}(T) = \eta_m^{-1} \Delta \Cp + \sfc_1 T$. Here,
$\eta_m$ is the \smf, $\Delta \Cp = C_{0\unit{T}} -
C_{12\unit{T}}$ is the difference in heat capacity between the
superconducting and normal state, and $\sfc_1$ is the Sommerfeld
coefficient of pure \fu. $\sfc_1$ turns out to be $\sfc_1 =
79\unit{\mu J/g/K^2}$ ($44\unit{mJ/mol_\mathrm{f.u.}/K^2}$) and
this results in $C_\mathrm{es}$ shown in
Fig.~\ref{fig:Ces_vs_TdivTc}. The sample shown in the figure has a
mass fraction of \fu\ of $\eta_m = 74.9\unit{\%}$, corresponding
to a volume fraction of $82.6\unit{\%}$. Further thermodynamic
parameters obtained from the measurements by CEA are listed in
Table \ref{tab:TD_params}.

After having determined the intrinsic Sommerfeld coefficient
$\sfc_1$ of \fu, it is possible to analyze the specific heat
anomaly at $\Tc$. The normalized specific heat jump $\Delta
\Cp\vert_{T_\mathrm{c}}/(\sfc_1 \Tc) = 1.9$ is significantly
larger than that for a weak coupling superconductor. It
corresponds to an electron-phonon coupling constant
$\lambda_\mathrm{ep} = 2 \int_0^\infty \alpha^2 F(\omega)/\omega
\, \mathrm{d}\omega \approx 1$,\cite{MaCoCa1987}, where $\alpha^2
F(\omega)$ is the electron-phonon spectral density. Thus, \fu\ is
a superconductor in the intermediate-coupling regime.

With the calculated band Sommerfeld coefficient $\sfc_\mathrm{b} =
17.8\unit{\mu J/g/K^2}$ of \fu\ from Ref.~\cite{KuJePi2004}, the
present result indicates a significant enhancement of the
electronic specific heat of
$(1+\lambda_\mathrm{ep}+\lambda_\mathrm{add}) = (79.1\unit{\mu
J/g/K^2})/(17.8\unit{\mu J/g/K^2}) \approx 4.4$. This enhancement
surpasses the one found in Sr$_2$RuO$_4$ of about $3.8$ by
$16\unit{\%}$ \cite{SaMeYeShFr2004}. Additional to the
electron-phonon enhancement $\lambda_\mathrm{ep}$, there is a
strong enhancement of unknown origin $\lambda_\mathrm{add} \approx
2.4$. We use the band structure $\sfc_\mathrm{b}$ from another
calculation to estimate the uncertainty in this additional
enhancement: Saniz \textit{et al.}~(Ref.~\cite{SaMeYeShFr2004})
have calculated the band Sommerfeld coefficient for KOs$_2$O$_6$,
which is $18\unit{\%}$ higher than the one calculated in
Ref.~\cite{KuJePi2004}. Assuming these $18\unit{\%}$ to be the
uncertainty in $\sfc_\mathrm{b}$ for \fu\ results in
$\lambda_\mathrm{add} \approx 2.1\pm0.3$. In view of a calculated
Stoner enhancement of the magnetic susceptibility of roughly
$2$,\cite{KuJePi2004} we speculate that the additional enhancement
is due to spin correlation effects.

\begin{table}
\caption{\label{tab:TD_params} Thermodynamic parameters of \fu. }
\begin{tabular}{lr}
Parameter & Value \\
\hline
$\xi, \lambda_\mathrm{eff}(0\unit{K})$ & $74\unit{\AA}, 252\unit{nm}$ \\
$\kappa(\Tc)$, $\kappa(0\unit{K})$ & $23$, $34$ \\
\hline
$\Delta \Cp\vert_{T_\mathrm{c}}/(\sfc_1 \Tc)$ & 1.9 \\
$\lambda_\mathrm{ep}, \lambda_\mathrm{add}$ & $1\pm0.1, 2.1\pm0.3$ \\
$b/\Tc$ & $(0.175\pm0.005)\unit{K^{-2}}$ \\
$\ce_1(0\unit{K})$ & $860\unit{\mu J/g}$ ($483\unit{mJ/mol_\mathrm{f.u.}}$) \\
\hline
$\Hc, H_\mathrm{c1}, H_\mathrm{c2}(0\unit{K})$ & $1249, 92, 60000\unit{Oe}$ \\
$-\mathrm{d}\Hc/\mathrm{d}T, -\mathrm{d}H_\mathrm{c2}/\mathrm{d}T \vert_{T_\mathrm{c}}$ & $369, 12000 \unit{Oe/K}$ \\
$Q \equiv - \frac{2\Tc}{\Hc(0)} \left.\frac{\mathrm{d}\Hc}{\mathrm{d}T}\right\vert_{T_\mathrm{c}} $ & $3.79\pm0.05$ \\
$k_\mathrm{B}\Tc/(\hbar\omega_\mathrm{ln})$ & $0.06$ \\
$2\Delta(0\unit{K})/(k_\mathrm{B}\Tc)$ & $3.87$\\
$1/(8\pi) \cdot (\sfc_1\Tc^2)/\ce_1$ & $0.15$\\
\hline
$\sfc_1$ & $79\unit{\mu J/g/K^2}$ ($44\unit{mJ/mol_\mathrm{f.u.}/K^2}$) \\
\end{tabular}
\footnotetext[1]{$b \cdot \Tc = Q^2/2$ has a universal value of
$5.99$ for a weak coupling BCS superconductor. This results in
$b/\Tc = 0.146\unit{K^{-2}}$ and $b = 0.94\unit{K^{-1}}$ for a BCS
superconductor with $\Tc=6.4\unit{K}$.}\footnotetext[2]{From
Ref.~\cite{BrKaZhKaBa2004}}
\end{table}

In summary, we apply the CEA to \fu\ to extract its intrinsic
thermodynamic properties. Among other parameters it allows us to
precisely determine the electron-phonon enhancement
$\lambda_\mathrm{ep} = 1\pm0.1$, i.e.~\fu\ is an
intermediate-coupling superconductor. Furthermore, \fu\ has a high
Sommerfeld coefficient for a pyrochlore of
$44\unit{mJ/mol_\mathrm{f.u.}/K^2}$ and thus a remarkably large
enhancement over the calculated band coefficient of $3.8$ to
$4.4$. In addition to the enhancement due to the coupling to
phonons, there is thus an unusually high enhancement of the
density of states due to other mechanisms $\lambda_\mathrm{add} =
2.1\pm0.3$. We speculate that the origin of this mass enhancement
lies in the $3$-dimensional triangular nature of the pyrochlore
lattice.

This study was partly supported by the Swiss National Science
Foundation.


\end{document}